\documentclass[letter]{aa}

\usepackage{natbib}

\usepackage{txfonts}

\usepackage[english]{babel}
\usepackage{graphicx}
\usepackage{verbatim}
\usepackage{multirow}
\usepackage{subfigure}
\usepackage{hyperref}
 \hypersetup{colorlinks=true,citecolor=blue}
\usepackage{longtable,lscape}

\begin{document}

\title{The contribution of spin to jet-disk coupling in black holes}
\author{Sjoert van Velzen\inst{1} \and Heino Falcke\inst{1,2,3}}
\institute{
   Department of Astrophysics/IMAPP, Radboud University, P.O. Box 9010, 6500 GL Nijmegen, The Netherlands \\\email{s.vanvelzen@astro.ru.nl}
   \and ASTRON, Dwingeloo, The Netherlands 
   \and Max-Planck-Institut f\"ur Radioastronomie Bonn, German
}

\authorrunning{van Velzen \& Falcke}
\titlerunning{The contribution of spin to jet-disk coupling}

\date{Received June 24 2013; accepted July 31 2013}

  \abstract
  { The spin of supermassive black holes could power jets from active galactic nuclei (AGN), although direct observational evidence for this conjecture is sparse. The accretion disk luminosity and jet power, on the other hand, have long been observed to follow a linear correlation. 
}
  {If jet power is coupled to black hole spin, deviations from the jet-disk correlation for  a sample of AGN can be used to probe the dispersion of the spin parameter ($a$) within this sample. }
 {To obtain a large sample of radio-loud AGN, we matched double-lobed radio sources identified in Faint Images of the Radio Sky at Twenty-centimeters (FIRST, 1.4~GHz) to spectroscopically confirmed quasars from the Sloan Digital Sky Survey (SDSS). We obtain 763 FR~II quasars with a median redshift of $z=1.1$.}
 { A tight correlation between the optical luminosity of the accretion disk and the lobe radio luminosity is observed. We estimate that 5--20\% of the bolometric disk luminosity is converted to jet power. Most of the scatter to the optical-radio correlation is due to environment; deviations from jet-disk coupling due to internal factors (e.g., spin) contribute at most 0.2~dex. 
}
 {Under the assumption that the Blandford-Znajek mechanism operates in AGN, we obtain an upper limit of 0.1~dex to the dispersion of the product of the spin and the magnetic flux threading the horizon. If black hole spin determines the AGN jet efficiency, then our observations imply that all FR~II quasars have very similar spin. In addition, the quasar spin distribution needs to have a wide gap to explain the radio-quiet population. 
The alternative, and perhaps more likely, interpretation of the tight jet-disk correlation is that black hole spin is not relevant for powering AGN jets.
}
\keywords{}
  \maketitle

\section{Introduction}
Accreting astrophysical objects often produce jets; outflows are commonly observed to escape from active black holes, neutron stars, white dwarfs, and young stellar objects. For all but the last, their velocities are (at least mildy) relativistic \citep{Reipurth01, Koerding08b, Mirabel99}. Despite a wealth of observations, the launch mechanism of jets has so far remained elusive.

Some authors argue that all disk-jet systems can be described in a scale-invariant framework \citep[e.g.,][]{Falcke99, McHardy06, Migliari06, Koerding06}. Jets from black holes, on the other hand, are often considered a separate class, since these are the only objects where General Relativity can have a major influence on the dynamics. In particular, it has often been conjectured that black hole spin is the dominant parameter in determining jet power, because space-time dragging can twist magnetic field lines \citep{BlandfordZnajek}. In the Blandford-Znajek mechanism, jet power is coupled to the dimensionless spin parameter ($a$) and the magnetic flux ($\Phi_B$) that threads the horizon: $Q_j\propto a^2\Phi_B^2$. 

Evidence for spin-powered jets has recently been claimed for stellar mass black holes in X-ray binaries. \citet{Narayan12} observed a correlation between the 5~GHz radio luminosity of the transient ballistic jet, which is launched during the accretion state transition when the accretion rate is close to the Eddington limit \citep{Fender04}, and the spin parameter of the black hole. \citet*{Fender10}, on the other hand, reported no evidence for a correlation between spin and a number of measurements that probe jet power.
The discrepancy between these two results (e.g., \citealt{Russell13} versus \citealt{Steiner13}) could be due to source selection.

The spin of supermassive black holes is often invoked to explain why only about 10\% of quasars are radio-loud \citep{Kellermann89, Xu99,Balokovic12}: only rapidly spinning black holes produce powerful jets \citep[e.g.,][]{Rees82, Wilson95, Moderski98, Sikora07}. In this is so-called spin-paradigm, 
the observed radio loudness maps to the supermassive black hole spin distribution. This distribution can be estimated by modeling black hole growth from accretion. If an accretion episode is long enough, maximum spin ($a\approx 1$) is expected because the inner parts of the accretion disk are forced to rotate in the equatorial plane of the black hole \citep{Bardeen75}. Simulations of spin evolution that include both accretion and black hole coalescence \citep{Hughes03, Gammie04} typically yield a smooth distribution of $a$; there is no gap between the high-spin ($a>0.9$) systems and the rest of the population \citep[e.g.,][]{Volonteri07}. 

Alternatively, the low number of radio-loud quasars can be explained by intermittent jet production due to state changes of the accretion disk \citep{Merloni03,Nipoti05}. Hot accretion or thick disks may be required to accumulate sufficient magnetic flux near the black hole horizon \citep{Beckwith08,Sikora13}.

A direct measurement of the spin of supermassive black holes can be obtained when the broad iron fluoressence line is detected at high signal-to-noise, with simultaneous broadband X-ray coverage to constrain line distortions due to absorption \citep{Risaliti13}.  Hence this is currently possible only for a restricted class of (nearby) AGN.

To study supermassive black hole spin for large populations, we propose to use the well-known correlation between jet power and disk luminosity. Jet-disk coupling has been observed using emission from the compact radio core \citep{Falcke95I}, the extended radio lobes \citep{Rawlings91}, and the power in X-ray cavities \citep{Allen06}; these observations can be used to show that a constant fraction of about 10\% of the disk luminosity is injected into the jet \citep{Falcke95II, Koerding08}.  

Since mass will disappear into the black hole while magnetic fields can pile up at the horizon, a linear coupling of accretion rate and jet power is not a natural result of the standard Bandford-Znajek mechanism. However, in the magnetically arrested disk \citep[MAD;][]{Narayan03} scenario, \label{page:MAD} black holes that are saturated with magnetic flux obey $Q_j = \eta_{\rm MAD} \dot{M}c^2$, with  $\eta_{\rm MAD}=140\%$ for $a=0.99$ \citep{Tchekhovskoy11}.

\begin{figure}
\begin{center}
\includegraphics[trim=10mm 0mm 0mm 4mm, clip, width=0.5\textwidth]{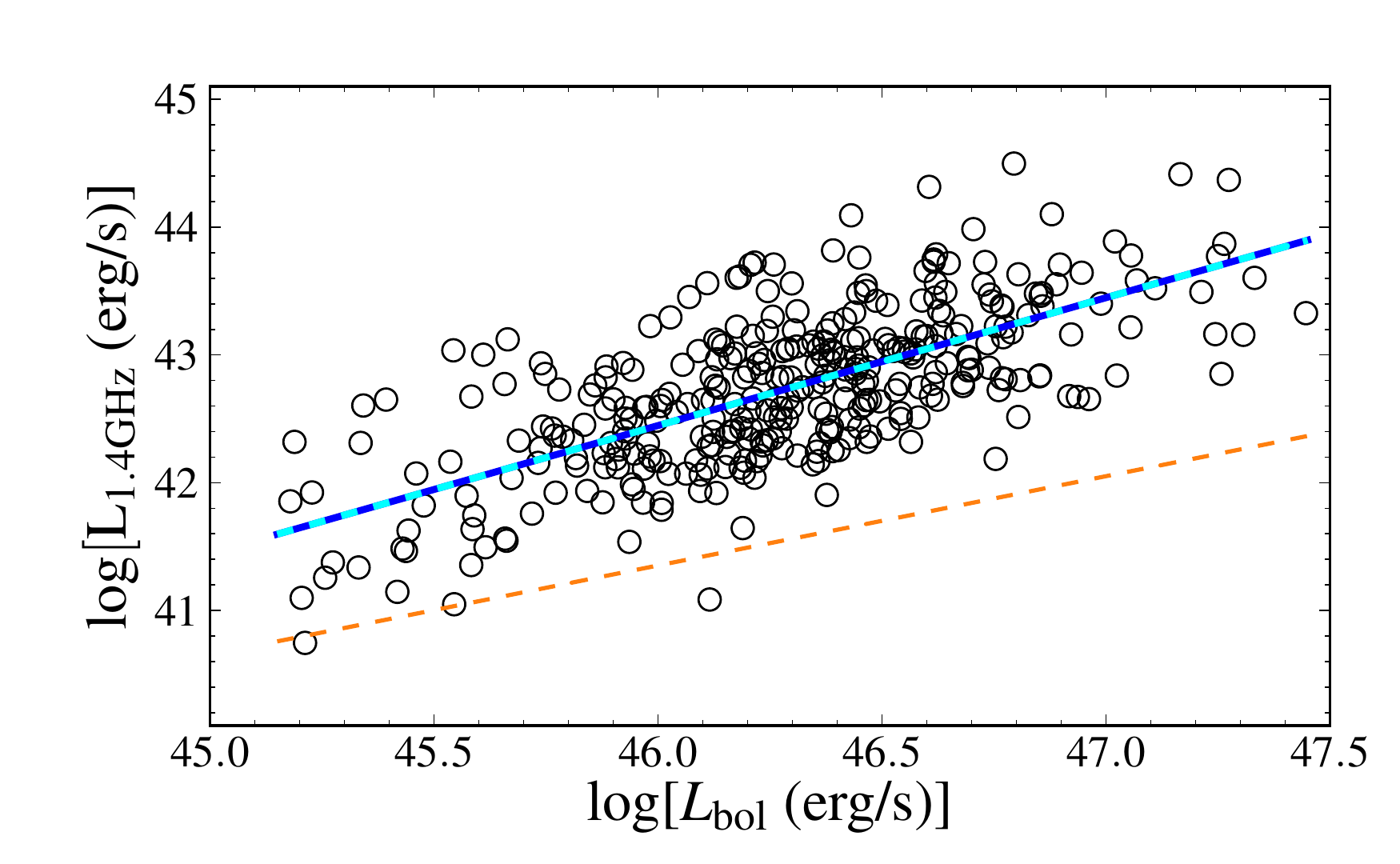}
\caption{We observe a linear correlation between the lobe radio luminosity and the bolometric quasar luminosity (solid blue+cyan line). The dashed orange line shows the correlation that one would obtain if all the radio sources clustered at the detection threshold (Eq.~\ref{eq:Smin}). Our observations are sensitive to outliers of the correlation of 1~dex.}\label{fig:Bol-Lradio}
\end{center}
\end{figure}

In this Letter, we will show that the scatter to the radio-optical correlation for a sample of AGN can be used to constrain the dispersion of black hole spin, or any other internal parameter that couples to jet power. 
Our approach is different from previous radio-based studies \citep[e.g.,][]{Nemmen07, Daly09}, which aimed to reproduce the observed jet power within the spin paradigm. Since Doppler boosting of the compact radio core of the jet smears out the radio-loudness distribution \citep{Falcke96}, we will only use the radio emission of the lobes of FR~II \citep{Fanaroff74} sources. These lobes emit isotropically and can thus be used to obtain an orientation-independent measurement of the jet power. 
If accretion onto compact objects is scale-invariant, radio-loud quasars can be considered \citep[e.g.,][]{Koerding06} to be the supermassive analogy of the ballistic jets that are launched when X-ray binaries switch between accretion states. 
Recalling that a correlation between radio luminosity and black hole spin has been reported only for these jets \citep{Narayan12}, FR~II quasars are arguably the most promising class of AGN to show the effects of spin on jet power.

We adopt the following cosmological parameters $h=0.70$, $\Omega_{\rm M}=0.3$, $\Omega_\Lambda=0.7$.

\begin{figure}
\begin{center}
\includegraphics[trim=10mm 0mm 0mm 4mm, clip, width=0.5\textwidth]{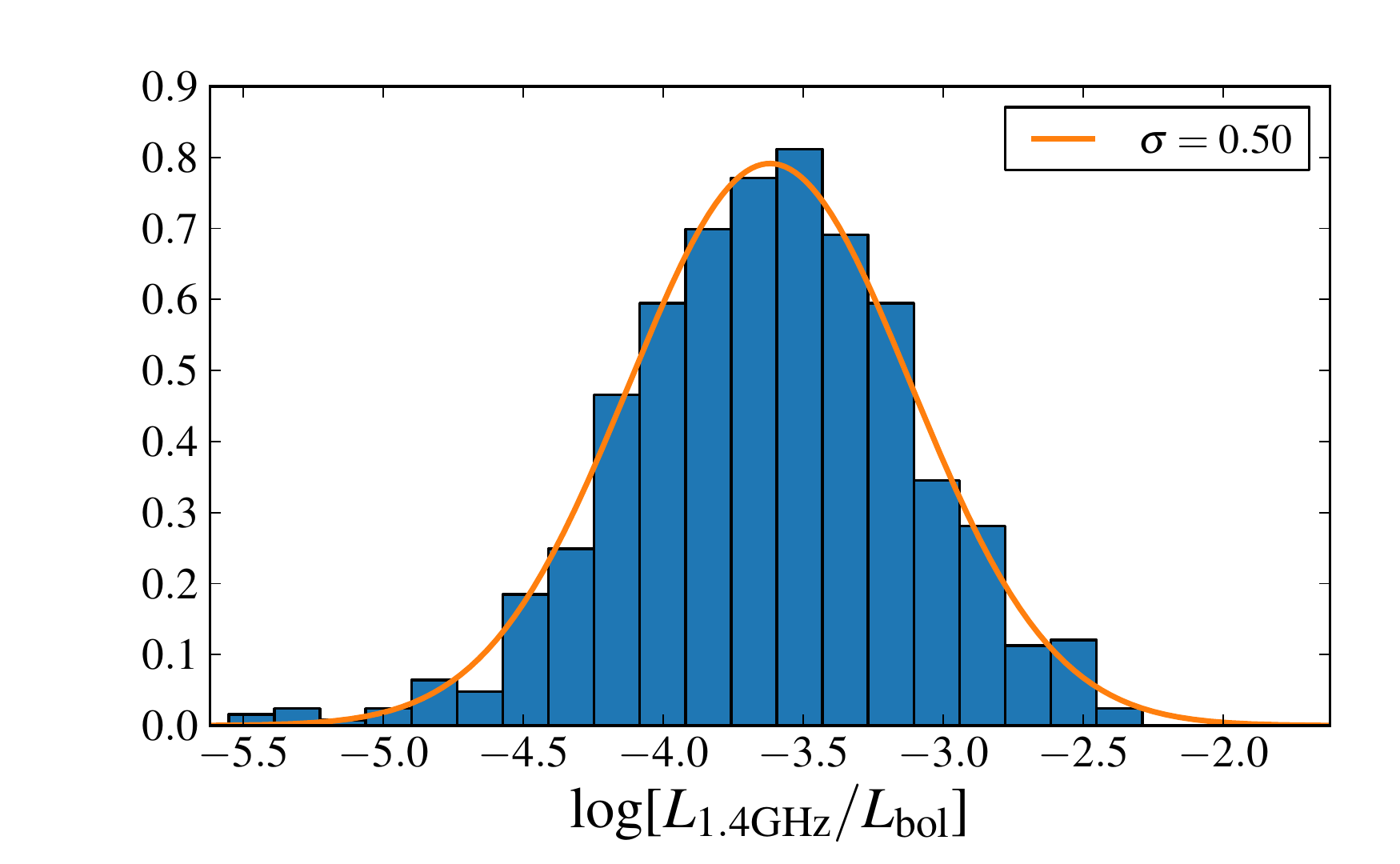}
\caption{The residuals of the correlation between bolometric disk luminosity and lobe radio luminosity (Fig.~\ref{fig:Bol-Lradio}) are well-described by a Gaussian distribution with $\sigma=0.50\pm0.01$~dex. }\label{fig:scatter}
\end{center}
\end{figure}

\section{Selection of FR~II quasars}\label{sec:data}
The radio emission of FR~II sources is dominated by two edge-brightened lobes that are separated by $\sim 10^2$~kpc. These sources are easy to identify in images of the radio sky; a surveys with a resolution a 10~arcsec can resolve pairs with a projected separation 100~kpc throughout the entire universe. 

In an accompanying paper (van Velzen \& Falcke, 2013, in prep) we demonstrate that a large sample of FR~IIs can be obtained by simply searching for source pairs in catalogs of radio surveys. We use the catalog radio sources at 1.4~GHz from FIRST \citep[Faint Images of the Radio Sky at Twenty-centimeters;][]{becker95}. These images have a resolution of $5$~arcsec FWHM and cover $\sim 10^4$~deg$^2$ to a limiting peak flux of 1~mJy. We selected all pairs within a separation $10"<d<60"$, yielding 89,214 candidate FR~IIs. 

To assess the potential of constraining black hole spin with radio observations, we have to consider the flux limit of our FR~II sample. To detect an extended source requires higher signal-to-noise than for a point source, so the limit on the integrated flux is a function of source compactness. Since we search for source pairs, our flux limit also depends on the lobe-lobe flux ratio. We thus obtain the following expression for the lower limit on the sum of the integrated flux of the two lobes 
\begin{equation}\label{eq:Smin}
S_{\nu, \rm lim} = 1~{\rm mJy} \times {\rm max}(S_{\rm int} / S_{\rm peak}) \times (1+f_{\rm lobe-lobe}) \quad.
\end{equation}
Here 
${\rm max}(S_{\rm int} / S_{\rm peak})$ is the largest value of the inverse compactness ratio of the two lobes (its median value is 1.8), and $f_{\rm lobe-lobe}$ denotes the absolute value of the lobe-lobe flux ratio. Since the lobe-lobe flux ratio peaks at unity, the median flux limit is 4~mJy. 

We matched the centers of our radio-selected FR~IIs to spectroscopically identified quasars \citep{Richards02, schneider07} in the Sloan Digital Sky Survey (SDSS). We use the Seventh Data Release \citep[DR7;][]{Abazajian09} edition of the catalog \citep{Schneider10}, consisting of 105,783 quasars with $M_i<-22$. For the match radius we use $d/4$, with $d$ the lobe-lobe separation (this ensures that the quasar is located between the two lobes). We obtain 763 FR~II quasars, the contribution of random matches is only $1.6\pm 0.5$\%, because the areal density of both the radio doubles and quasars is relatively sparse. The median redshift is $z=1.1$; the median mass and Eddington ratio of the FR~II quasars is $10^{9.2}~M_\odot$ and 8\%, respectively \citep{Shen11}.

Next, we use the luminosity and frequency dependent bolometric correction of \citet*{Hopkins07} to convert the $i$-band flux to the bolometric quasar luminosity ($L_{\rm bol}$). The $K$~correction for the rest-frame radio luminosity is obtained using the mean spectral index of the lobes ($\alpha=-0.9$). We obtain a linear relation between bolometric disk luminosity and the 1.4~GHz (rest-frame) lobe luminosity (Fig.~\ref{fig:Bol-Lradio}) with $\left< {\rm log}\, L_{\rm bol} / L_{\rm 1.4 GHz} \right>=-3.6$. The residuals of this correlation follow a normal distribution with $\sigma_{\rm optical-radio}=0.50\pm0.01$~dex, as obtained using an unbinned maximum likelihood fit (Fig.~\ref{fig:scatter}). We also observed a highly significant flux-flux correlation (the p-value for a Kendall's tau rank analysis is $10^{-33}$), so our optical-radio correlation is not induced by Malmquist bias. 

\begin{figure}
\begin{center}
\includegraphics[trim=10mm 0mm 0mm 4mm, clip, width=0.5\textwidth]{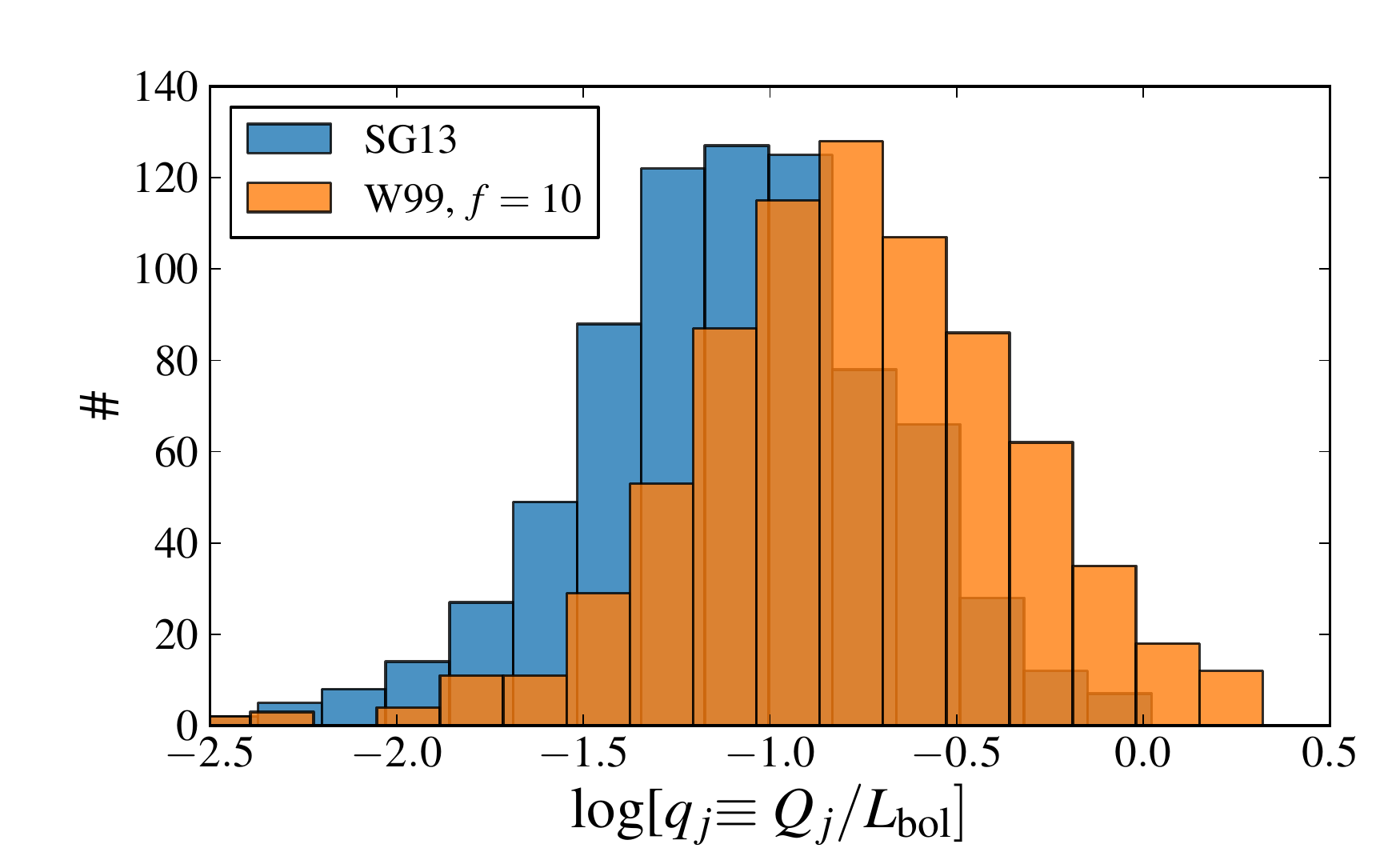}
\caption{The jet-disk coupling parameter for two different models of synchrotron emission by the lobes. The \citet[][W99]{Willott99} model is based on minimum energy arguments; the median jet power in this model is $q_j=0.2$. For the \citet[][SG13]{Shabala13} scaling we obtain $q_j\approx 0.1$. }\label{fig:qj_hist}
\end{center}
\end{figure}

The observed optical-radio correlation is consistent with the \citet{Willott99} minimal energy argument for synchrotron emission, $Q_j \propto L_{\rm bol}^{6/7} f^{3/2}$. For a fudge factor that is typical for FR~II sources, $f=10$ \citep{Blundell00}, we find that on average about 20\% of the bolometric quasar luminosity ends up in the jet, $q_j\equiv Q_j / L_{\rm bol}  = 0.17$. Recently, \citet{Godfrey13} presented a method to measure jet power from the hotspot properties. These authors obtain $Q_j \propto L^{0.8} D^{0.58}$, with $D$ the source size \citep{Shabala13}. For this model, the median jet-disk coupling parameter is $q_j=0.08$ (Fig.~\ref{fig:qj_hist}).

\begin{figure}
\begin{center}
\includegraphics[trim=10mm 0mm 0mm 4mm, clip, width=0.5\textwidth]{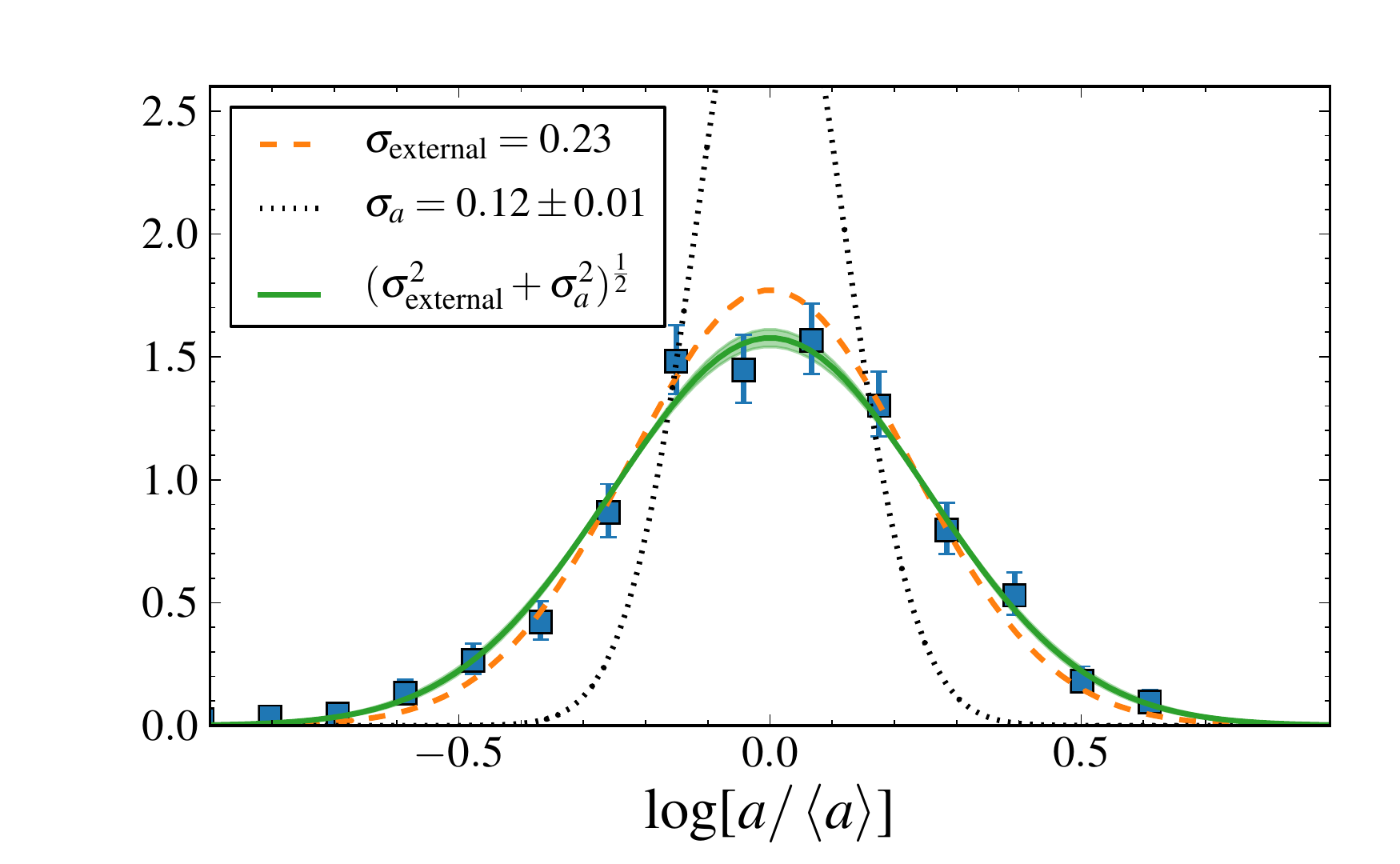}
\caption{The dispersion of black hole spin as obtained from the residuals to the radio-optical correlation (Fig.~\ref{fig:scatter}). Here we adopt the Blandford-Znajek mechanism to change coordinates, $L_{\rm 1.4GHz}\propto Q_{j} \propto \Phi_B^2  a^2$. The lower limit on the external factors (dashed orange line) provides a reasonable description of the data. Hence the contribution due to spin or magnetic flux is constrained to a narrow range (black dotted line). 
}\label{fig:scatter-a}
\end{center}
\end{figure}

\section{Measuring the spin contribution}\label{sec:results}
In the previous section we obtained a sample of 763 FR~II quasars and found a linear relation between the lobe radio luminosity and the bolometric disk luminosity. The residuals to this correlation can be modeled with four separated Gaussian distributions: 
\begin{equation}
\sigma_{\rm radio-optical}^2 =\sigma_{\rm data}^2 + \sigma_{\rm environ}^2 + \sigma^2_{\rm delay}+ \sigma_{\rm internal}^2 = 0.50~{\rm dex}.
\end{equation}
Below we present a lower limit for the first three external effects, allowing us to obtain an upper limit to $\sigma_{\rm internal}$.

The scatter due to the measurement uncertainty ($\sigma_{\rm data}$) is dominated by the dispersion in the bolometric correction that we applied to the optical luminosity. For quasars observed at optical wavelengths, this dispersion is measured to be 0.15~dex \citep{Hopkins07}. 

From symmetry arguments, we can assume that two jets from one black hole carry equal kinetic energy, hence the lobe-lobe flux ratio can be used to obtain a lower limit on the effect of the local environmental on differences in the radio luminosity ($\sigma_{\rm environ}$). The lobe-lobe flux ratio can be modeled with a Gaussian distribution with $\sigma=0.39$~dex. Under the conservative assumption that the quasar-to-quasar fluctuations in environment are similar to the fluctuations probed by two jets from one quasar, we obtain $\sigma_{\rm environ}>0.39$~dex. 

The radio emission is delayed and smeared with respect to our optical view of the quasar, since it takes time for the energy produced near the black hole to reach the lobes. Evolution of the quasars during this $\sim 10^6$~yr delay will also introduce scatter to the optical-radio correlation. 
The longest baselines of optical quasar light curves reach 50~yr (in the observer's frame); a structure function analysis reveals 0.28~mag variability for this time scale \citep{Macleod12}. We thus obtain the very conservative lower limit, $\sigma_{\rm delay}>0.11$~dex. 
We note that for quasars with a radio luminosity that is a factor ten below the radio-optical correlation (Fig.~\ref{fig:Bol-Lradio}), we find a median core-to-lobe ratio of 1.4. Only 6\% of all FR~II quasars in our sample have such strong cores, providing evidence that the largest outliers to the optical-radio correlation are due to a delay between the newly activated disk luminosity and the lobe radio luminosity. 

The three measurements discussed above can be combined to find a strict lower limit on the external contributions to deviations from jet-disk coupling, $\sigma_{\rm external} > 0.45$~dex. 
We now obtain an upper limit for the remaining contribution by processes that could influence the jet power, $\sigma_{\rm internal}=0.24\pm 0.03$~dex. Adopting the standard Blandford-Znajek mechanism, this translates into an upper limit on the dispersion of the product of black holes spin and magnetic flux, $\sigma_{\rm BZ}<0.12$~dex (Fig.~\ref{fig:scatter-a}).

\section{Discussion}\label{sec:disc}
We observed a tight correlation between radio and optical luminosity using a large sample of FR II quasars at $z\sim 1$. After correcting for the environment and other external factors that contribute to the scatter around this correlation, we find that factors that are internal to the black hole can change the jet power by at most 0.24~dex. For jets powered by a Blandford-Znajek mechanism, $Q_j \propto \Phi_B^2 a^2$, this implies $\sigma_{\rm BZ}<0.12$. 

Most of the SDSS quasars are not detected as FR~II radio sources in the FIRST survey \citep[see also][]{deVries06, White07}. 
Adopting the spin paradigm, this would imply that the spin of the radio-quiet quasars is at least a factor of three lower than the mean spin of the FR II quasars. The typical value for this spin-reduced jet power is $2\times10^{38}$~W; this is much larger than the median FR II jet power \citep[e.g.,][]{Antognini12}, hence such radio-quiet quasars would not be missing from our sample because their jets are too weak to obtain an FR~II morphology. We also note that the spin-alignment time of quasars \citep[e.g.,][]{Volonteri07} is at least an order of magnitude longer than the typical lifetime of powerful FR~IIs \citep[][]{Carilli91}, so our radio selection is not biased to high spin black holes. 

Our results appear to be at odds with the analysis of X-ray binaries by \citet{Narayan12}, who observed a correlation between radio luminosity and black hole spin that spans three orders of magnitude. If radio-loud quasars also follow this correlation, the small scatter to the optical-radio correlation implies that these supermassive black holes all have a similar spin parameter. We consider a narrow distribution of black hole spin for radio-loud quasars, plus a wide gap to explain the radio-quiet quasars, an unnatural result of black hole evolution. Since spin-up and spin-down are both expected to occur during the growth of a black hole, one generically obtains a smooth distribution for the spin parameter. 

For quasars at $z\sim 1$, massive accretion events triggered by major mergers may yield a high ($a>0.6$) mean spin, with modest scatter \citep{Volonteri12}. In this scenario, which is consistent with our upper limit on the dispersion of spin for FR~IIs, jet-spin coupling via the Blandford-Znajek mechanism clearly fails to explain why the majority of quasars are radio-quiet. To explain our observations within the spin paradigm, one is forced to adopt a very strong jet-spin coupling, such that high spin effectively acts as a threshold for jet production \citep[][]{Tchekhovskoy10}. In this case, however, one would expect an asymmetric distribution of residuals, which is not observed (Fig.~\ref{fig:scatter}). 

In the MAD scenario, introduced on page~\pageref{page:MAD}, jet power is independent of magnetic flux and is proportional to the accretion rate. Hence for $a=1$ this model would naturally predict linear jet-disk coupling with no scatter due to internal parameters. Unfortunately, the predicted MAD jet efficiency ($\eta_{\rm MAD} \approx 1.4 \, a^2$) for maximally spinning black holes is a factor of 100 higher than the observed mean efficiency of the FR~II quasars (Fig.~\ref{fig:qj_hist}): $Q_j \approx 0.1 L_{\rm bol} \approx 10^{-2} \dot{M}c^2$.

We thus conclude that the observed radio-optical correlation supports a tight jet-disk coupling \citep{Rawlings91,Falcke95I} and suggest that the power output from the inner part of the accretion disk dominates over the power extracted from the black hole by the Blandford-Znajek mechanism \citep{Livio99}. Spin, or in fact any other internal parameter that can vary from quasar to quasar, is not a dominant parameter for powering jets. This leaves intermittent jet production due to state changes of the accretion disk as the best explanation for the radio loudness distribution of quasars. 


{\it Acknowledgments } ---  We would like the thank the referee, Demos Kazanas, for the swift reply. We are grateful to P.\,L. Biermann, J. Dexter, T.\,M. Heckman, J. Hlavacek-Larrondo, E. K\"ording, J.\,H. Krolik, J.\,E. McClintock, B.\,D. Metzger, R. Narayan, R.\,S. Nemmen, H. R\"ottering, B.\,S. Sathyaprakash, M. Sikora,  J.\,F. Steiner, R.\,L. White, and N.\,L. Zakamska for useful discussions. 

This work was supported by an ERC Advanced Grant (n. 227610). The National Radio Astronomy Observatory is a facility of the National Science Foundation operated under cooperative agreement by Associated Universities, Inc. Funding for the SDSS and SDSS-II has been provided by the Alfred P. Sloan Foundation, the Participating Institutions, the National Science Foundation, the U.S. Department of Energy, the National Aeronautics and Space Administration, the Japanese Monbukagakusho, the Max Planck Society, and the Higher Education Funding Council for England. 

\bibliography{general_desk}

\end{document}